\documentclass[prb,aps,twocolumn,amsmath,superscriptaddress]{revtex4}

\usepackage{upgreek}
\usepackage{graphicx}
\usepackage{epsfig}
\usepackage{textcomp,gensymb} 
\usepackage{array}
\newcolumntype{P}[1]{>{\centering\arraybackslash}p{#1}}

\begin{document}
\title{Carrier Diffusion in Thin-Film CH$_3$NH$_3$PbI$_3$ Perovskite Measured using Four-Wave Mixing}

\author{D. Webber}
\affiliation{Department of Physics and Atmospheric Science,
Dalhousie University, Halifax, Nova Scotia B3H 4R2 Canada}

\author{C. Clegg}

\affiliation{Department of Physics and Atmospheric Science,
Dalhousie University, Halifax, Nova Scotia B3H 4R2 Canada}

\author{A. W. Mason}

\affiliation{Department of Physics and Atmospheric Science,
Dalhousie University, Halifax, Nova Scotia B3H 4R2 Canada}

\author{S. A. March}

\affiliation{Department of Physics and Atmospheric Science,
Dalhousie University, Halifax, Nova Scotia B3H 4R2 Canada}

\author{I. G. Hill}

\affiliation{Department of Physics and Atmospheric Science,
Dalhousie University, Halifax, Nova Scotia B3H 4R2 Canada}

\author{K. C. Hall}

\affiliation{Department of Physics and Atmospheric Science,
Dalhousie University, Halifax, Nova Scotia B3H 4R2 Canada}

\begin{abstract}
We report the application of femtosecond four-wave mixing (FWM) to the study of carrier transport in solution-processed CH$_3$NH$_3$PbI$_3$. The diffusion coefficient was extracted through direct detection of the lateral diffusion of carriers utilizing the transient grating technique, coupled with simultaneous measurement of decay kinetics exploiting the versatility of the boxcar excitation beam geometry.  The observation of exponential decay of the transient grating versus interpulse delay indicates diffusive transport with negligible trapping within the first nanosecond following excitation.  The in-plane transport geometry in our experiments enabled the diffusion length to be compared directly with the grain size, indicating that carriers move across multiple grain boundaries prior to recombination. Our experiments illustrate the broad utility of FWM spectroscopy for rapid characterization of macroscopic film transport properties. 
\end{abstract}

\maketitle
Organo-lead trihalide perovskites possess a unique combination of electronic and optical properties, making them attractive for applications in light-emitting diodes,\cite{Tan:2014,Kim:2015} lasers,\cite{Deschler:2014,Zhu:2015,Fu:2016} optical sensors,\cite{Dou:2014,Fang:2015,DongAdv:2015} and most notably high performance solar cells where the efficiencies have rapidly increased, reaching over 22\% in just a few years.\cite{NREL:2016} In addition, these hybrid perovskites can be solution-processed and applied as thin films to a variety of substrates,\cite{Docampo:2013,Li:2016} pointing to the potential for large-scale, low-cost solar cell production.\cite{Barrows:2014,JHKim:2015,Razza:2015,Deng:2015,Hambsch:2016} The transport properties of electrons and holes within the perovskite absorber material are crucial to the performance of photovoltaics and other optoelectronic technologies using these materials.  The first observation of micron-scale carrier diffusion lengths in CH$_3$NH$_3$PbI$_{3-x}$Cl$_x$\cite{Stranks:2013,Xing:2013,Stoumpos:2013} stimulated a comprehensive research effort aimed at understanding the nature of carrier transport.\cite{Wehrenfennig:2014,GonzalezPedro:2014,Li:2015,Guo:2015,Hutter:2015,Chen:2015,Liu:2016,Edri:2014,Ponseca:2014,Valverde:2015,DongSci:2015,Shi:2015,Adhyaksa:2016,Semonin:2016} Carrier mobilities and/or diffusion lengths have been studied using electrical techniques (\textit{e.g.} AC Photo Hall,\cite{Chen:2015} space-charge-limited current,\cite{Shi:2015} impedance spectroscopy,\cite{GonzalezPedro:2014}, and spatially-resolved electron-beam-induced current\cite{Edri:2014}) as well as optical techniques that rely on electrical contacts such as photoluminescence quenching\cite{Stranks:2013,Xing:2013} and scanning photocurrent microscopy.\cite{Semonin:2016,Liu:2016}  Some of these techniques offer the ability to probe transport in a working solar cell device, however imperfectly characterized interface energetics and ambiguities tied to the model-dependent extraction of transport characteristics impede the determination of the physical processes limiting device performance.

\begin{figure}[ht]\vspace{0pt}
    \includegraphics[width=8cm]{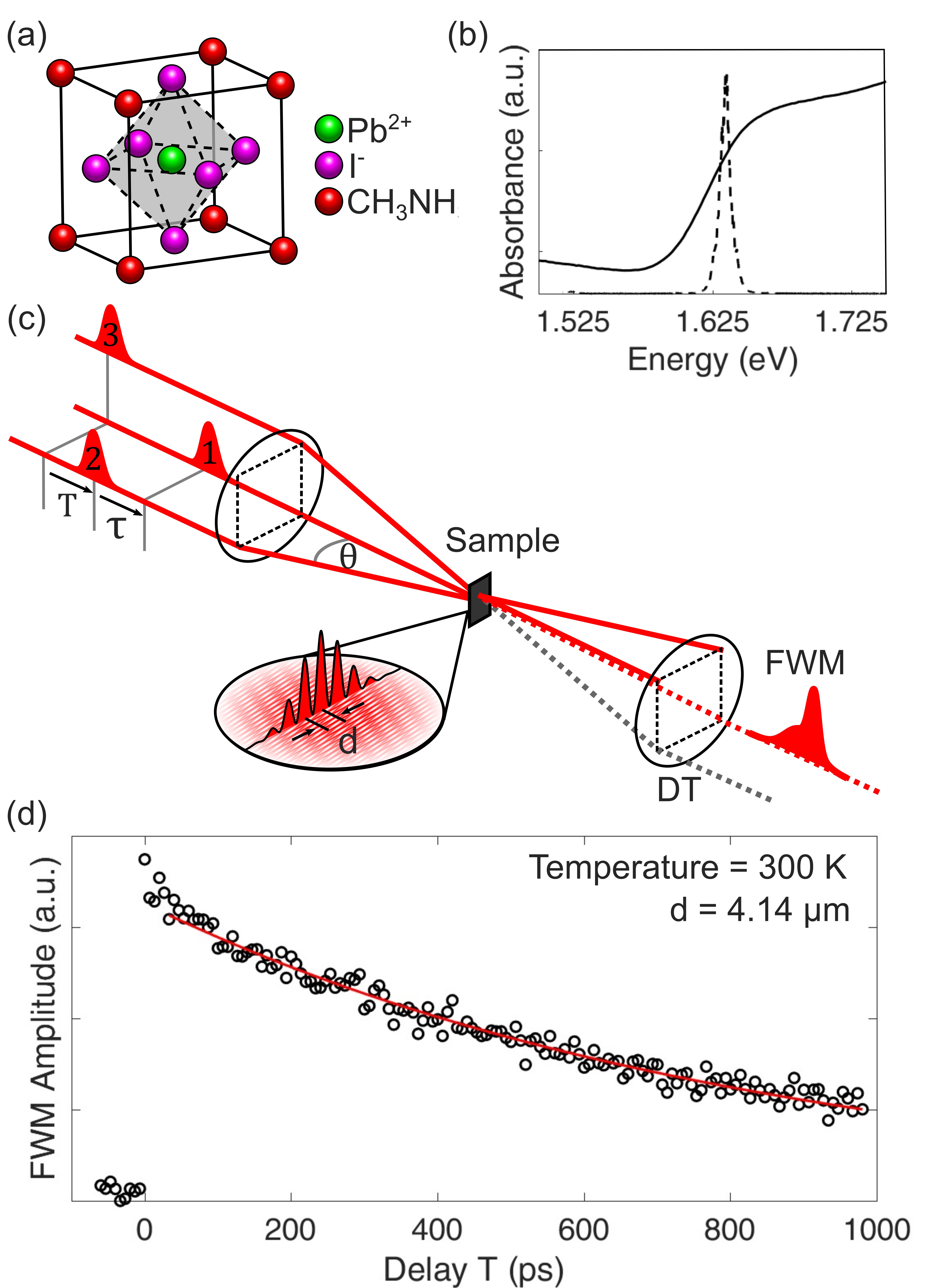}
    \caption{(color online) (a) The crystal structure of CH$_3$NH$_3$PbI$_3$. (b) The linear absorption spectrum for the pervoskite sample at 300~K (solid curve) alongside the laser pulse spectrum (dashed curve). (c) A schematic of the FWM setup showing the boxcar geometry.  (d) FWM signal as a function of delay T at 300~K for d = 4.14 $\upmu$m. The solid curve is a least-squares fit to Eqn.~\ref{eqn:expdecay}.}
    \label{fig:Figure1}
\end{figure}

All-optical techniques provide an effective approach to studying carrier transport within a wide range of photovoltaic materials as no carrier extraction layers or ohmic contacts are required.  Time-domain terahertz spectroscopy (TDTHz) and time-resolved microwave conductivity (TRMC) have provided valuable insight into carrier scattering processes in the organometal halide perovskites in recent years.\cite{Ponseca:2014,Valverde:2015,Hutter:2015} For such techniques, the quantitative determination of mobility requires modeling of the Drude response of the multi-component carrier system.  Transient absorption microscopy\cite{Guo:2015} provides a more direct, all-optical probe of carrier transport.  In this technique, carrier motion away from the laser excitation spot is imaged as a function of time after excitation using a diffraction-limited focusing geometry, enabling the influence of variations in sample morphology on the local transport characteristics to be assessed.  A complementary all-optical technique that probes carrier transport on macroscopic length scales, while also relying on the direct detection of the spatial motion of carriers, would provide a valuable characterization tool for the rapid survey of photovoltaic materials.

\begin{figure}[htb]\vspace{0pt}
    \includegraphics[width=8.5cm]{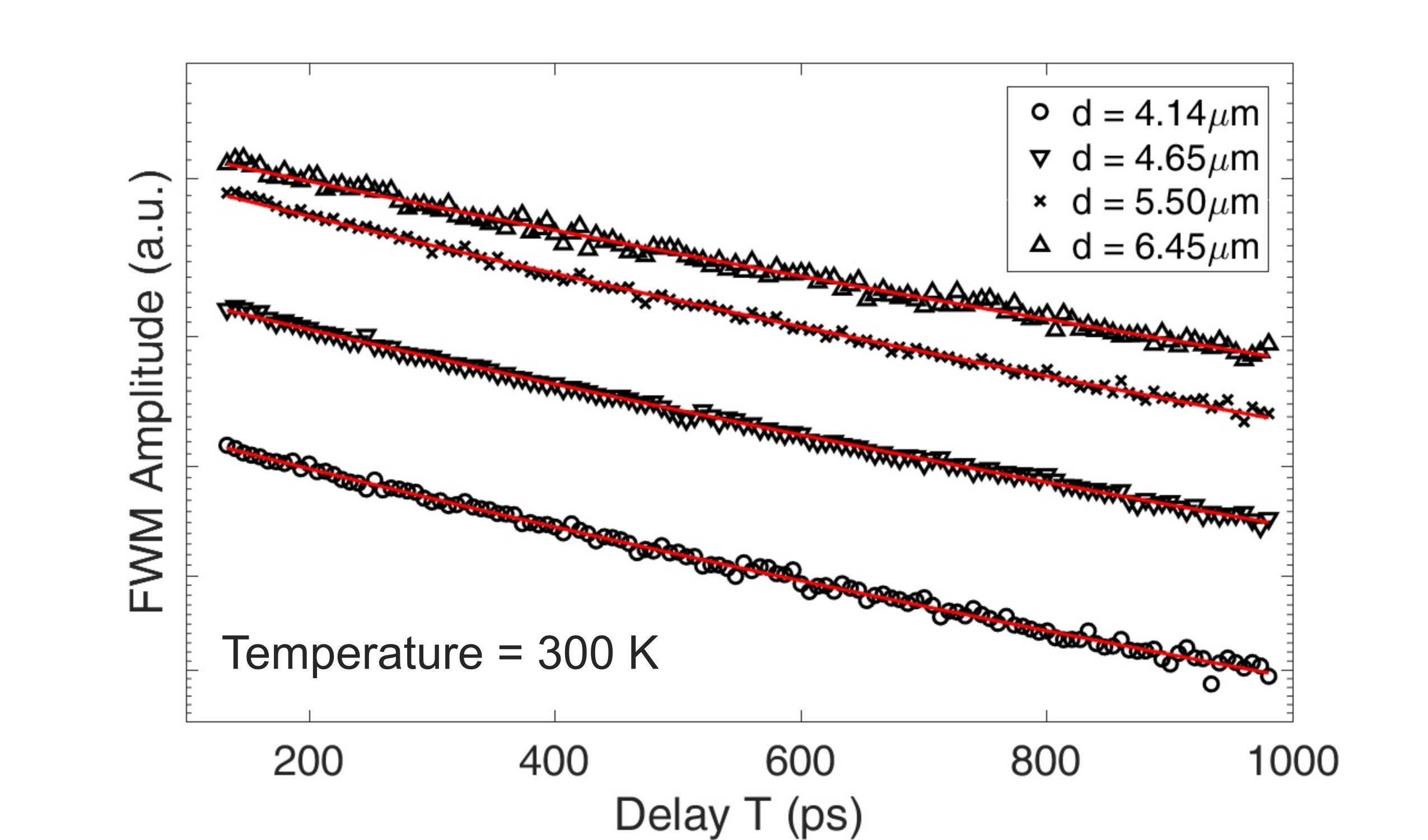}
    \caption{(color online)  FWM signal intensity on a logarithmic scale as a function of delay T for a range of grating constants. The sample temperature is 300~K. The solid curves are fits using Eqn.~\ref{eqn:expdecay}.}
    \label{fig:Figure2}
\end{figure}

Here we apply femtosecond four-wave mixing (FWM) to investigate carrier transport in CH$_3$NH$_3$PbI$_3$, highlighting the utility of this technique for application to a wide range of organic-inorganic perovskites.  Our experiments illustrate several advantages of this approach: (i) it involves a direct measurement of the ambipolar diffusion coefficient without any need for modeling; (ii) it is an all-optical (contactless) technique, applicable to the survey of a wide range of photovoltaic materials; (iii) the carrier lifetime may be accessed through fitting the grating period dependence or directly through differential transmission with the same apparatus, exploiting the versatility of the boxcar geometry; (iv) diffusive and trap-limited transport are easily distinguished through the interpulse delay dependence of the four-wave mixing signal; and (v) the in-plane transport configuration enables the motion of carriers to be directly compared with grain sizes and allows anisotropy of transport characteristics\cite{Liu:2016} to be evaluated through simple rotation of the sample.  As an additional feature, transport may be studied using this technique with high time resolution at low carrier densities ($\sim$10$^{15}$ to 10$^{16}$~cm$^{-3}$), reflective of solar cell operating conditions.  Our experimental results on solution-processed CH$_3$NH$_3$PbI$_3$ indicate diffusive transport within the first nanosecond following excitation, with a diffusion coefficient (\textrm{D}) of 1.7~$\pm$~0.1~cm$^2$~s$^{-1}$ and diffusion length (L$_{\textrm{D}}$) of 0.95~$\pm$~0.07~$\mu$m.  As this technique probes the ambipolar diffusion length, which is dominated by the transport of the slower carrier type (electrons or holes), the observation of L$_{\textrm{D}}\sim$four times larger than the measured average grain size of 250~nm indicates that defects tied to grain boundaries have a limited impact on transport, in agreement with recent experiments\cite{Chen:2015,Liu:2016} and theoretical predictions of predominantly shallow traps.\cite{YinAdvMater:2014,YinAPL:2014}   The application of nonlinear optical FWM techniques to the study of transport in CH$_3$NH$_3$PbI$_3$ presented here builds upon recent studies of carrier-carrier interactions and exciton localization in this material using a two-pulse degenerate FWM configuration,\cite{March:2016, MarchTwo:2016} as well as measurements of the third-order nonlinear susceptibility in perovskite thin films.\cite{Johnson:2016}   

In the four-wave mixing experiments, three noncollinear optical pulses with electric fields $\vec{\textrm{E}}_1$(t), $\vec{\textrm{E}}_2$(t-$\uptau$), and $\vec{\textrm{E}}_3$(t-($\uptau$+T)) with corresponding wavevectors $\vec{\textrm{k}_1}$, $\vec{\textrm{k}_2}$, and $\vec{\textrm{k}_3}$ were focused onto the sample in a noncollinear boxcar geometry,\cite{Bristow:2009} resulting in a fourth field emitted in the phase-matched direction $\vec{\textrm{k}_3} + \vec{\textrm{k}_2} - \vec{\textrm{k}_1}$ (See Fig.~\ref{fig:Figure1}(c)).  Interference between pulses $\vec{\textrm{E}}_1$(t) and $\vec{\textrm{E}}_2$(t-$\uptau$) creates a periodic spatial modulation of the carrier density in the material, forming a transient grating.  The grating undergoes decay due to recombination and lateral carrier transport in the material. A third delayed pulse $\vec{\textrm{E}}_3$(t-($\uptau$+T)), was used to probe the grating modulation depth by monitoring the diffracted signal as a function of time delay T between pulses $\vec{\textrm{E}}_2$ and $\vec{\textrm{E}}_3$. Provided carrier transport is diffusive,\cite{Salcedo:1978,Ruzicka:2010} the FWM signal decays exponentially versus T,\cite{Salcedo:1978,Schwab:1992} with the signal intensity proportional to:
\begin{equation}
\textrm{I(T)} \propto \exp{\left(-\textrm{T}/\textrm{T}_\textrm{s}\right)}.
\label{eqn:expdecay}
\end{equation}
and the decay constant T$_{\textrm{s}}$ determined by:
\begin{equation}
\frac{1}{\textrm{T}_\textrm{s}} = \frac{2}{\textrm{T}_\textrm{L}} + \frac{8 \pi^2 \textrm{D}}{\textrm{d}^2}, 
\label{eqn:eqn1}
\end{equation}
In Eqn.~\ref{eqn:eqn1}, T$_{\textrm{L}}$ is the carrier lifetime, D is the ambipolar diffusion coefficient, $\lambda$ is the center wavelength of the laser pulse, and the grating constant $\textrm{d} = \frac{\lambda}{2\sin{\left(\uptheta/2\right)}}$, where $\uptheta$ is the angle between the two incident beams.  The slope of a linear fit of the inverse grating decay time versus $\frac{8 \pi^2}{\textrm{d}^2}$ yields the ambipolar diffusion coefficient D, while the offset yields the carrier lifetime.  The experimental boxcar geometry also permits differential transmission (DT) measurements using the same apparatus for a separate verification of the carrier lifetime.  In this case, $\vec{\textrm{E}}_1$(t) was blocked, $\vec{\textrm{E}}_2$(t-$\uptau$) served as the pump pulse (setting $\uptau$~=~0), and $\vec{\textrm{E}}_3$(t-T) attenuated by approximately 100$\times$ served as the probe pulse.  

\begin{figure}[htb]\vspace{0pt}
    \includegraphics[width=8.5cm]{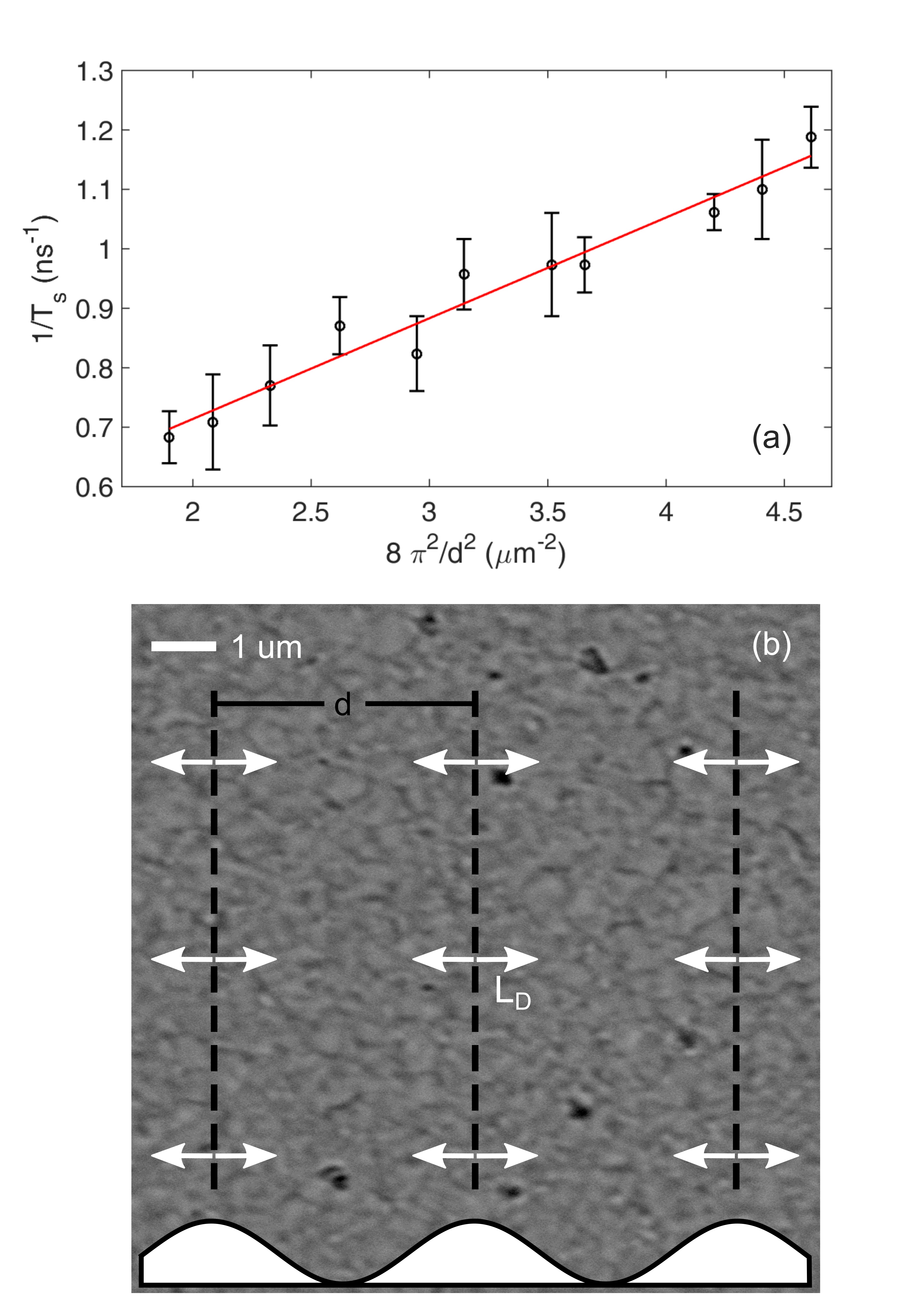}
    \caption{(color online)  (a) Inverse signal decay time T$_{\textrm{s}}^{-1}$ as a function of $8 \pi^2/\textrm{d}^2$ measured at 300~K. The solid line is a fit of the data to Eqn.~\ref{eqn:eqn1}, yielding the parameters D = 1.7~cm$^2$~s$^{-1}$ and T$_{\textrm{L}}$ = 5.3~ns. (b) Scanning electron microscopy image of the CH$_3$NH$_3$PbI$_3$ sample.  The initial spatial variation of the carrier distribution for a grating period of d = 4.14 $\mu$m is represented by the white filled region at the bottom, and the white arrows (with a magnitude equal to the measured diffusion length) represent diffusive carrier motion driven by the concentration gradient.}
    \label{fig:Figure3}
\end{figure} 

The laser source used in these experiments was a Ti:sapphire oscillator producing 100~fs pulses with a 10~nm bandwidth.  The grating period d was varied by controlling the angle $\uptheta$ between the beams $\vec{\textrm{E}}_1$(t) and $\vec{\textrm{E}}_2$(t-$\uptau$) by changing the effective focal length of a two-lens system. The excitation density in the sample was held constant as a function of d by measuring the beam diameter (150 - 200~$\upmu$m) for each lens configuration and adjusting the laser power using neutral density filters to maintain a constant pulse fluence.  All three optical pulses were s-polarized.  At each delay T, the FWM signal as a function of the delay $\uptau$ between $\vec{\textrm{E}}_1$(t) and $\vec{\textrm{E}}_2$(t-$\uptau$) was detected with a fast amplified photodiode using a rapid scan technique mediated by a retroreflector attached to a vibrating speaker cone.  This allowed the FWM signal at $\uptau = 0$~fs to be easily separated from scattered light sources propagating along the same direction. The delay T between $\vec{\textrm{E}}_2$(t-$\uptau$) and $\vec{\textrm{E}}_3$(t-($\uptau$+T)) was varied using a computer-controlled delay stage.   

The CH$_3$NH$_3$PbI$_3$ sample under investigation in this work was prepared using a modified sequential deposition procedure described in detail in Ref. \onlinecite{March:2016}.   Scanning electron microscopy (SEM) measurements showed good uniformity, and x-ray diffraction (XRD) indicated full conversion to perovskite with no residual lead iodide.  For the XRD and SEM studies, the exposure of the sample to air was limited to 1-2 hours.  The linear absorption spectrum of a companion sample prepared at the same time using an identical procedure was measured with a Cary UV-Vis spectrometer, and is shown alongside the laser spectrum in Fig.~\ref{fig:Figure1}(b).  For all FWM experiments, exposure to air was avoided by mounting the sample in a compact optical cryostat within an argon glove box and sealing prior to transport to the FWM apparatus.  

\begin{figure}[htb]\vspace{0pt}
    \includegraphics[width=8.5cm]{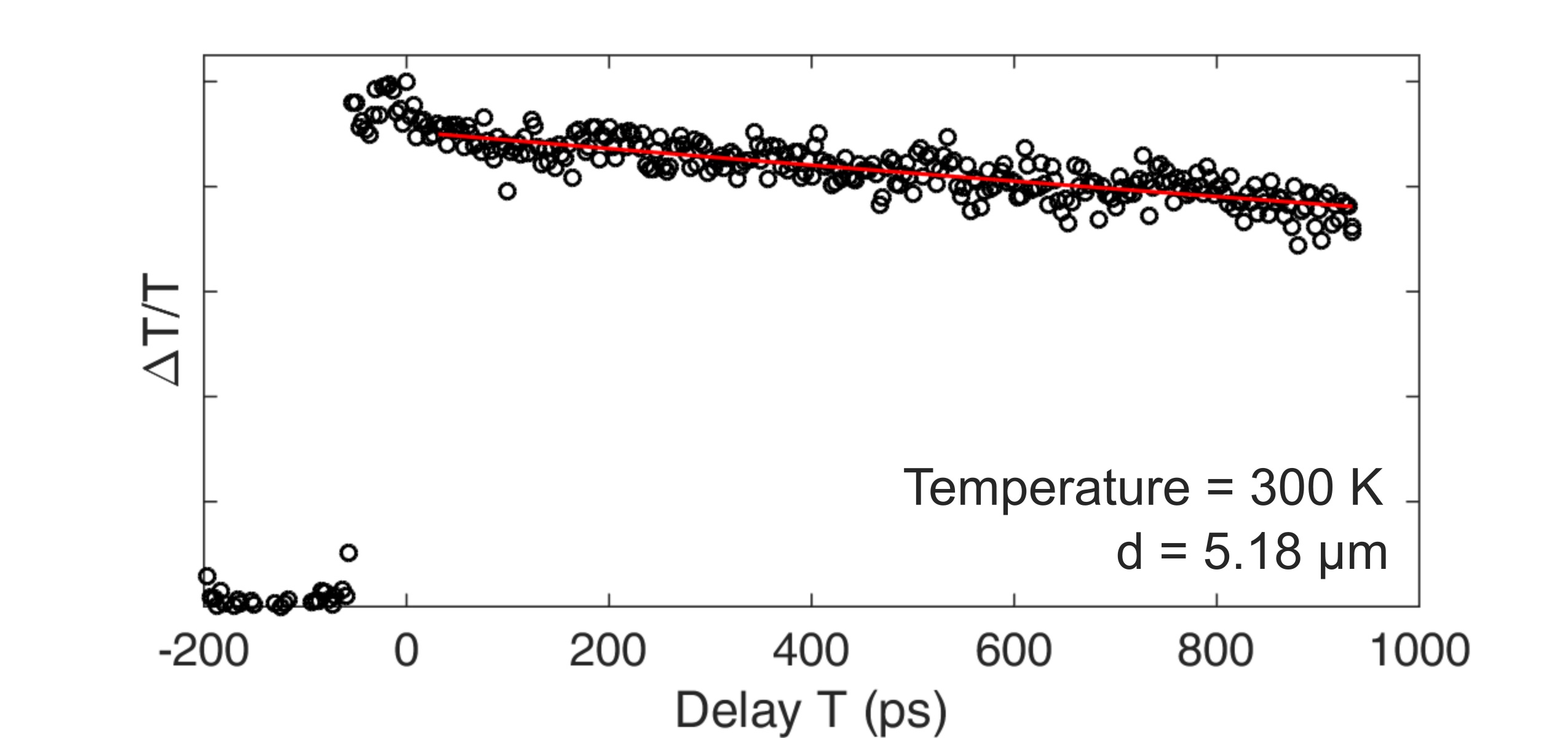}
    \caption{(color online)  Results of differential transmission measurements using the same FWM boxcar apparatus on the CH$_3$NH$_3$PbI$_3$ sample under the same experimental conditions. The solid curve shows the result of a least-squares fit of the measured signal to an exponential decay, yielding T$_{\textrm{L}}$~=~5.4~ns.}
    \label{fig:Figure4}
\end{figure}

Fig.~\ref{fig:Figure1}(d) shows the FWM signal at $\uptau$~=~0~fs versus pulse delay T at 300~K for a grating period of d~=~4.14~$\upmu$m. For these experiments, the laser was tuned to 760 nm, exciting carriers $\sim$10-20~meV above the band gap\cite{Even:2014} and the excitation density was 1.6$\times$10$^{16}$~cm$^{-3}$.  With the exception of the region of overlap of the excitation laser pulses, the four-wave mixing signal exhibits exponential decay over the full measurement range of 1 ns.  The red curve in Fig.~\ref{fig:Figure1}(d) shows a fit to Eqn.~\ref{eqn:expdecay}, yielding a grating decay time of T$_{\textrm{s}}$~=~840 ps.  Fig.~\ref{fig:Figure2} shows a logarithmic plot of the FWM signal as a function of delay for grating constants d = 4.14, 4.65, 5.50, and 6.45 $\upmu$m. As shown by the quality of the linear fits, the FWM signal decays exponentially in all cases. The inverse decay time T$_{\textrm{s}}^{-1}$ is plotted versus $8\pi^2/\textrm{d}^2$ in Fig.~\ref{fig:Figure3}.   The error bars represent the standard deviation of ten delay scans taken for each grating constant.  The solid line in Fig.~\ref{fig:Figure3} is a linear fit using Eqn.~\ref{eqn:eqn1}. The slope (intercept) yields the ambipolar diffusion coefficient (carrier lifetime), corresponding to \textrm{D}~=~1.7~$\pm$~0.1~cm$^2$~s$^{-1}$  and T$_{\textrm{L}}$~=~5.3~$\pm$~0.5~ns. As a self-consistency check, differential transmission measurements were also carried out using the same apparatus for the same total excitation carrier density (Fig.~\ref{fig:Figure4}).  These measurements yield a carrier lifetime of $\textrm{T}_{\textrm{L}}$ = 5.4~$\pm$~0.2~ns, in good agreement with the value of 5.3~ns obtained from fitting the transient grating results.  Using a one dimensional diffusion model, for which L$_{\textrm{D}}$ = $\sqrt{\textrm{D}\textrm{T}_{\textrm{L}}}$, these fit results yield a diffusion length of 0.95~$\pm$~0.07~$\mu$m. This value is in line with measurements of the ambipolar diffusion length on similar films using transient absorption microscopy techniques.\cite{Guo:2015}

The good agreement between the experimental results and Eqn.~\ref{eqn:expdecay} and ~\ref{eqn:eqn1} over the full range of accessible delays indicates that carrier transport is within the diffusive regime.   Carrier trapping, which would lead to subdiffusive behavior,\cite{Akselrod:2014} therefore has a negligible influence on the carrier kinetics within the first nanosecond after excitation.  This is in agreement with studies of CH$_3$NH$_3$PbI$_3$ thin films using transient absorption microscopy,\cite{Guo:2015} in which a diffusive model accounted for the spatiotemporal carrier occupation over a 3 ns time interval after excitation.  Using broadband THz spectroscopy, Valverde-Chavez \textit{et al.} observed a rapid initial decay of mobility on a subpicosecond time scale in single crystal CH$_3$NH$_3$PbI$_3$ attributed to the charging of trap states and/or a charge-mediated phonon scattering mechanism.\cite{Valverde:2015}  The experiments in Ref.~\onlinecite{Valverde:2015} were carried out at a much larger excited carrier density ($\sim$10$^{18}$ cm$^{-3}$), suggesting that these charge-mediated effects turn on under stronger excitation conditions than our experiments.  

Scanning electron microscopy measurements on the same CH$_3$NH$_3$PbI$_3$ film used for FWM experiments are shown in Fig.~\ref{fig:Figure3}(b).  A spatial Fourier analysis indicates the presence of grains up to 500~nm in diameter, with a mean grain size of 250~nm.  The observation of L$_D$ $\sim$4 $\times$ larger than the mean grain diameter, together with the fact that this ambipolar diffusion length is limited by the slower carrier species,\cite{Ruzicka:2010} indicates that grain boundaries do not constitute a significant barrier to carrier motion.  This result is in agreement with recent measurements on solution processed CH$_3$NH$_3$PbI$_3$ using AC Photo Hall\cite{Chen:2015} and photocurrent imaging techniques\cite{Liu:2016} as well as theoretical calculations suggesting that the most prevalent point defects represent shallow traps and that grain boundaries have a relatively benign effect.\cite{YinAdvMater:2014,YinAPL:2014}   In addition, screening by the methylammonium ions may play a role in reducing the rate of trapping by charged defects,\cite{Chen:2015} in line with the recently proposed large polaron model of carrier transport in perovskites.\cite{XYZ}  

Reported values of the diffusion length in solution processed methylammonium lead iodide films have varied over the range 0.1 to 23~$\mu$m.\cite{Stranks:2013,Xing:2013,Wehrenfennig:2014,GonzalezPedro:2014,Li:2015,Guo:2015,Hutter:2015,Chen:2015,Liu:2016,Edri:2014}  Differences in both the microstructure of the film and the optically-excited carrier density contribute to this broad range of values as these together dictate the carrier lifetime.\cite{Semonin:2016}  The application of a variety of experimental techniques to extract the transport characteristics also plays an important role in the observed variability.  For instance, photoluminescence-based techniques only probe the transport of radiative species, TRMC and TRTHz techniques measure the sum of the electron and hole mobilities ($\mu_e$+$\mu_h$), AC Photo Hall measurements detect $\mu_h$-$\mu_e$, and TA microscopy and the FWM technique applied in this work both probe the ambipolar diffusion coefficient ($\mu_a = \frac{2\mu_e\mu_h}{\mu_e + \mu_h}$).  In addition, while techniques that rely on electrical contacts can offer the advantage of probing transport in a working device,\cite{Edri:2014} uncertainties tied to the energetics of the contact interfaces can significantly influence the extracted results for L$_{\textrm{D}}$.\cite{Hodes:2015}  This highlights the need for a survey tool for probing transport in a wide range of photovoltaic materials such as the FWM transient grating technique presented here.  We note that a steady-state photocarrier grating method relying on a change of photocurrent induced by interference of continuous wave laser beams\cite{Ritter:1986,Balberg:1988} was recently applied to organic-inorganic perovskite films.\cite{Adhyaksa:2016}  In contrast to the approach used in Ref.~\onlinecite{Adhyaksa:2016}, the transient grating technique applied in this work does not require electrical contacts to the sample and provides access to dynamic information about the photocarrier response.

In conclusion, we have applied the all-optical four-wave mixing transient grating technique using a boxcar geometry to the study of carrier transport in a solution processed film of CH$_3$NH$_3$PbI$_3$, illustrating the power and flexibilty of this approach for application to emerging photovoltaic materials based on the organic-inorganic family of perovskites.  Our experiments yield values for the carrier lifetime of 5.3~$\pm$~0.5~ns and the ambipolar diffusion coefficient of 1.7~$\pm$~0.1~cm$^2$~s$^{-1}$, corresponding to a diffusion length of 0.95~$\pm$~0.07~$\mu$m.  These fit values are obtained from the measured transient grating signal without the need for modeling, a consequence of the direct optical detection of the spatial motion of carriers.  Our experiments show that this method provides a valuable rapid survey tool for probing macroscopic transport properties in the hybrid perovskites. 

This research is supported by the Natural Sciences and Engineering Research Council
of Canada, the Canada Foundation for Innovation, and the Canada Research Chairs program.

\end{document}